\documentclass[%
reprint,
superscriptaddress,
 amsmath,amssymb,
 aps,
 prl,
]{revtex4-2}

\usepackage[utf8]{inputenc}
\usepackage[margin=0.75in]{geometry}
\usepackage{graphicx}
\usepackage{bbold}
\usepackage{xcolor}

\everymath{\displaystyle}

\usepackage{physics}
\usepackage{enumitem}
\begin{document}
\title{Controlling Rydberg excitations using ion core transitions in alkaline earth atom tweezer arrays}
%\title{Coherent control of Rydberg excitations using ion core transitions in an alkaline earth atom}
\author{Alex P. Burgers}
\affiliation{Princeton University, Department of Electrical and Computer Engineering, Princeton, New Jersey 08544}
\author{Shuo Ma}
\affiliation{Princeton University, Department of Electrical and Computer Engineering, Princeton, New Jersey 08544}
\affiliation{Princeton University, Department of Physics, Princeton, New Jersey 08544}
\author{Sam Saskin}
\affiliation{Princeton University, Department of Electrical and Computer Engineering, Princeton, New Jersey 08544}
\affiliation{Princeton University, Department of Physics, Princeton, New Jersey 08544}
\author{Jack Wilson}
\affiliation{Princeton University, Department of Electrical and Computer Engineering, Princeton, New Jersey 08544}
\author{Miguel A. Alarc\'on}
\author{Chris H. Greene}
\affiliation{Purdue University, Department of Physics and Astronomy, West Lafayette, Indiana 47907}
\affiliation{Purdue University, Purdue Quantum Science and Engineering Institute, West Lafayette, Indiana 47907}
\author{Jeff D. Thompson}
\email[ ]{jdthompson@princeton.edu}
\affiliation{Princeton University, Department of Electrical and Computer Engineering, Princeton, New Jersey 08544}

% \date{May 2020}
\begin{abstract}
    Scalable, local control over gate operations is an outstanding challenge in the field of quantum computing and programmable quantum simulation with Rydberg atom arrays. One approach is to use a global field to excite atoms to the Rydberg state, and tune individual atoms in and out of resonance via local light shifts. In this work, we point out that photon scattering errors from light shifts can be significantly reduced if the light shift is applied to the Rydberg state instead of the ground state, which can be realized in Rydberg states of alkaline earth atoms using optical transitions in the ion core. As a proof-of-concept, we experimentally demonstrate global control of Rydberg excitations in a Yb optical tweezer array via light shifts induced by a laser tuned near the Yb$^+$ $6s\rightarrow6p_{1/2}$ transition. We also perform detailed spectroscopy of the induced light shift and scattering rates of the $6sns$ $^3$S$_1$ Rydberg states and reveal the existence of satellite lines where losses from autoionization are strongly suppressed. This work can be readily extended to implement local gate operations in Rydberg atom arrays.   
\end{abstract}

\maketitle
Neutral atoms trapped in reconfigurable optical tweezer arrays are a leading platform for quantum computation and quantum simulation. Key features of this approach are bottom-up control afforded by optical tweezer technology \cite{schlosser2001,endres2016,barredo2016} and strong, controllable interactions via Rydberg excitations \cite{lukin2001,saffman2010}. In recent years, this platform has been used to microscopically probe the properties and dynamics of quantum phase transitions \cite{bernien2017, deleseleuc2019, scholl2020, ebadi2020}, to generate and probe large-scale entangled states \cite{omran2019,choi2021}, and implement high-fidelity gate operations in multi-qubit arrays \cite{levine2019,graham2019,Madjarov2020}.

An outstanding challenge in the field of Rydberg atom arrays is scalable, local addressing of gate operations, especially for multi-qubit gates involving excitation to the Rydberg state. Generating rapidly switchable, reconfigurable and focused Rydberg excitation beams across an atomic array is demanding for optical modulators, as the intensity, pointing and frequency of these beams must be tightly regulated \cite{monroe2013}. This challenge is exacerbated for the wavelengths needed for single-photon Rydberg excitation (near or below 300 nm \cite{schauss2012,hankin2014,Madjarov2020}), for which very few optical materials are transparent. 

\begin{figure*}[]
\centering
\includegraphics[width=1\linewidth]{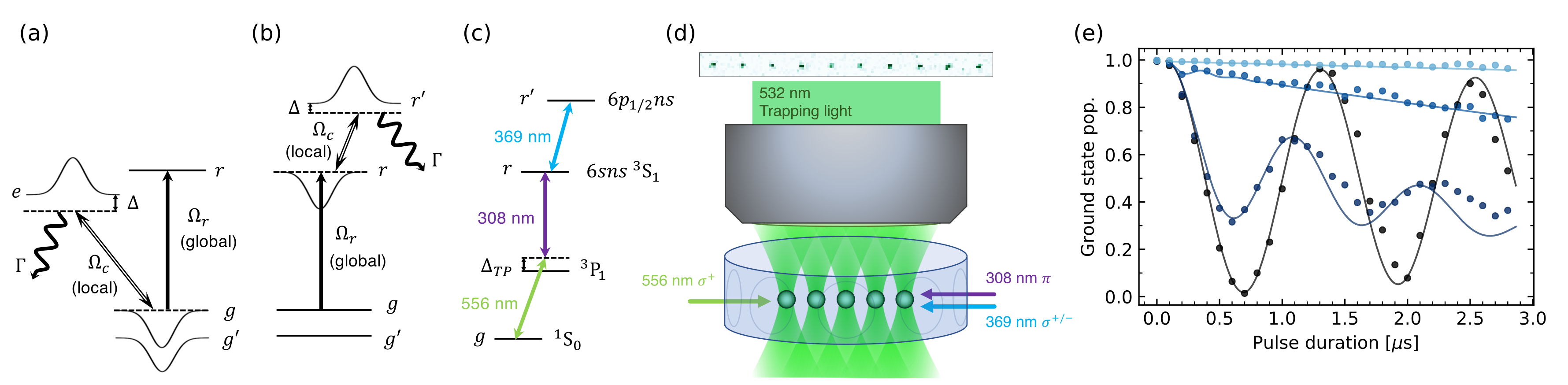}
\vspace{-0.5cm}
\caption{\label{fig:fig1}(a) In the conventional approach to local control with light shifts, the $\ket{g}\rightarrow\ket{r}$ transition is detuned via a shift on $\ket{g}$ generated by driving the $\ket{g}\rightarrow\ket{e}$ transition, using a control field with Rabi frequency $\Omega_c$ and detuning $\Delta$. $\ket{e}$ is an excited state with linewidth $\Gamma$. A second ground state $\ket{g'}$ used to encode a qubit will typically also experience a light shift. (b) In the approach presented here, the light shift is generated with a control field tuned near a transition from $\ket{r}$ to another highly excited state, $\ket{r'}$, which leaves the ground states unshifted. (c) Levels in $^{174}$Yb relevant to this work: $\ket{g}=$ $^1$S$_0$ is coupled to the Rydberg state $\ket{r}=$ $6sns$ $^3$S$_1$ by a two-photon transition through $^3$P$_1$ with intermediate detuning $\Delta_{TP}=40$ MHz. The control field drives the $6sns \rightarrow 6p_{1/2}ns$ transition. (d) Schematic of the experiment, indicating the propagation direction and polarization of the driving fields shown in (c). The Rydberg and control beams are focused to approximately 15 $\mu$m $1/e^2$ radius at the position of the atoms. (e) Rabi oscillations between $\ket{g}$ and $\ket{r}$ (black) are suppressed by the addition of increasing control beam intensities (black to blue: $I_c=$ 0, 15, 90, 600 W/cm$^2$). The solid lines show the result of a numerical simulation.}
\vspace{0mm}
\end{figure*}

An alternative approach is to apply local light shifts using non-resonant control beams, to tune the ground to Rydberg ($\ket{g} \rightarrow \ket{r}$) transition on individual sites out of resonance with a global Rydberg excitation beam (Fig. \ref{fig:fig1}a). In addition to relaxing the stability requirements for the addressing beam, this approach also readily scales to parallel gate implementation \cite{levine2019}, an important consideration for fault-tolerant quantum computing \cite{preskill1997}. Local control with light shifts has been demonstrated with Rydberg tweezer arrays \cite{labuhn2014, levine2018, omran2019}, and with microwave transitions in optical lattices \cite{weitenberg2011,wang2015}. In the particular case of implementing Rydberg blockade gates on qubits encoded in hyperfine atomic levels, the local light shift determines which atoms participate in the gate, and which are spectators \cite{labuhn2014}. To avoid spurious interactions with spectator atoms, the magnitude of the light shift must be much larger than the global Rydberg Rabi frequency. At the same time, photon scattering from local addressing beams must be suppressed to avoid gate errors. Consequently, to achieve high-fidelity operation these conditions must be balanced, requiring both a large detuning and large intensity of the control beam.

In this Letter, we point out that the scattering error can be significantly reduced if the control beam only couples to the Rydberg state, $\ket{r}$ (Fig. \ref{fig:fig1}b). In this case, the scattering errors occur primarily from $\ket{r}$, but the population of this state is suppressed for atoms illuminated by the control beam, which remain in the ground state. As a consequence, the control beam can be operated at much smaller detunings and lower powers.
%, and the intensity needed to achieve a given error rate scales as $I \propto 1/\epsilon^2$.
%This may result in a $10^4$-fold power reduction for gates with $\epsilon = 10^{-4}$.

Unfortunately, in alkali atoms such as Rb and Cs, there are no strong optical transitions connected to $\ket{r}$.
%, and its polarizability is dominated by the weak ponderomotive shift of the Rydberg electron \cite{dutta2000,zhang2011,li2013}.
In alkaline earth atoms (AEAs) with two valence electrons, the situation is different: the Rydberg electron orbits an optically active ion core, which has strong, allowed transitions. The far off-resonance polarizability of these transitions has been used for trapping Yb \cite{Wilson2019} and Sr \cite{cortinas2020} Rydberg atoms, while the short lifetime (via autoionization decay) of core excited states has been used for efficient state detection in Sr atomic gases \cite{Lochead2013} and tweezer arrays \cite{Madjarov2020}.

In this work, we explore the use of these ion core transitions to control Rydberg excitations in Yb (Fig. \ref{fig:fig1}). We first quantify the control field intensity $I_c$ needed to realize locally addressed operations with an error rate $\epsilon$ by shifting the ground states or the Rydberg states, and find that scaling is improved from $I_c \propto 1/\epsilon^3$ to $I_c \propto 1/\epsilon^2$ in the latter case. We then experimentally demonstrate control of Rydberg excitations in an optical tweezer array of $^{174}$Yb atoms \cite{Saskin2019,Wilson2019} using a light shift induced on the Rydberg state by a control beam tuned near the 369 nm $6s \rightarrow 6p_{1/2}$ transition in the Yb$^+$ ion core. We investigate the influence of the control beam on the $6sns$ $^3$S$_1$ Rydberg states in detail, and find that, near resonance, the light shift and autoionization scattering rate can be described by a two-level system model, consistent with the isolated core electron (ICE) approximation for doubly excited Rydberg states \cite{cooke1978}. We then use the global control beam to switch \emph{on} and \emph{off} the excitation of both single atoms and high-fidelity Bell states ($\mathcal{F} > 0.948(12)$) in a dimerized array, using a control beam intensity equivalent to only 1.3 $\mu$W in a diffraction-limited spot. Lastly, we observe that for larger control beam detunings (comparable to the spacing between Rydberg levels), satellite features called ``shake-up" lines \cite{gallagher1987} appear, giving rise to an additional suppression of the scattering rate without altering the light shift. We demonstrate an additional reduction in scattering errors from the control beam at these special detunings.

These proof-of-concept experiments demonstrate global control over Rydberg excitations from $\ket{g}\rightarrow\ket{r}$ and can be readily extended to locally controlled quantum gate operations. This extension requires two additional steps: focusing the 369 nm light down to single sites, and introducing a second stable level $\ket{g'}$ to define a qubit. This could be the $^3$P$_0$ clock state in $^{174}$Yb, or an additional hyperfine state in an odd isotope such as $^{171}$Yb. The latter approach has the advantage that the differential light shift on the qubit levels is extremely small, further relaxing the intensity stability requirements for the control beam.

We first establish the theoretical scaling of the addressing fidelity with the intensity of the control light. We consider the following scenario: given a regular array of atoms with qubits encoded in hyperfine ground states (Fig. 1a), we wish to apply an entangling Rydberg blockade gate using a global Rydberg beam \cite{levine2019,Saffman2020} (with Rabi frequency $\Omega_r$) to a sparse subset of nearest-neighbor qubit pairs, while realizing no operation (up to single-qubit phases) on the remaining, spectator qubits. Given a control beam that shifts the state $\ket{g}$ in the spectator atoms relative to $\ket{r}$ by an amount $\Delta_{LS} = \Omega_c^2/(4 \Delta)$, and induces a scattering rate $\Gamma_{LS} = \Gamma \Omega_c^2/(4 \Delta^2)$ \footnote{This is the total scattering rate including both Raman and Rayleigh scattering, but we treat it as Raman scattering when $\Delta < \Delta_f$, the fine structure splitting \cite{cline1994}. Given the required value of $\Delta$ derived below, this is self-consistent for gate errors $\epsilon \gtrsim 10^{-4}$ in Rb and Cs, given $\Delta_r/\Gamma \approx 10^6$ in these atoms.} (see Fig. 1a for variable definitions), there are two dominant errors. The first is the probability for a spectator atom to be excited off-resonantly to $\ket{r}$, $P_r \propto (\Omega_r/\Delta_{LS})^2$, and thereby blockade the intended gate, which results in an error probability $\epsilon_{rot} \propto P_r$. The second is a photon scattering error in the spectator atom, with a probability $\epsilon_{sc} \propto \Gamma_{LS} t_g$ ($t_g \propto 2\pi/\Omega_r$ is the gate duration). We note that the former error is improved with larger control field intensity, while the latter gets worse, resulting in a minimum total error that can only be reduced by increasing $\Delta$.

There is also an intrinsic gate error on the qubits participating in the gate, resulting from the finite lifetime of the Rydberg state, $\epsilon \propto t_g \Gamma_r = 2\pi \Gamma_r/\Omega_r$, where $\Gamma_r$ is the decay rate of $\ket{r}$. A natural condition is that the addressing errors should be comparable to the intrinsic error, $\epsilon_{rot} + \epsilon_{sc} = \epsilon$, and the minimum detuning and control intensity needed to realize this are $\tilde{\Delta} \propto \Gamma/\epsilon^{3/2}$, and  $\tilde{I}_c \propto |\tilde{\Omega}_c|^2 \propto \Gamma \Gamma_r/\epsilon^3$.

If the light shift is applied on the state $\ket{r}$ by coupling to another state $\ket{r'}$ (Fig. \ref{fig:fig1}b), the spectator qubit error probability $\epsilon_{rot}$ is the same, but now the scattering error is strongly suppressed, since only the $\ket{r}$ state experiences loss. In this case, $\epsilon_{sc} \propto \Gamma_{LS}t_g P_r$, with $P_r \propto \epsilon_{rot}$ as before. Since $\epsilon_{rot} \propto \Omega_c^{-4}$ and $\Gamma_{LS} \propto \Omega_c^{-2}$, both errors now decrease monotonically with increasing control power. The condition $\epsilon_{rot} + \epsilon_{sc} = \epsilon$ can be satisfied with $\tilde{\Delta} \propto \Gamma/\sqrt{\epsilon}$, and $\tilde{I}_c \propto |\tilde{\Omega}_c|^2 \propto \Gamma \Gamma_r/\epsilon^2$.

We now turn to an experimental demonstration of controlling Rydberg excitations by shifting $\ket{r}$ using an ion core transition. Our experiment begins by creating a 1D, defect-free array \cite{endres2016,barredo2016} of 10 $^{174}$Yb atoms in the $^1$S$_0$ ground state $\ket{g}$. This state is coupled to the $6s75s\,^3$S$_1$ ($m_J = +1$) Rydberg state $\ket{r}$ in a two-photon process via the $^3$P$_1$ state (Fig. 1c), with an intermediate state detuning of $\Delta_{TP}=40$ MHz and two-photon Rabi frequency $\Omega_r = 2\pi \times 0.7$ MHz. The 369 nm control beam with intensity $I_{c}$ is co-propagating with the Rydberg lasers, and is stabilized near the $\ket{r}\rightarrow \ket{r'}$ transition using a wavemeter with 60 MHz accuracy. At the end of the experiment, any population in $\ket{r}$ is removed using a second pulse of the 369 nm beam, at full power, before detecting the remaining atoms in $\ket{g}$ \cite{Madjarov2020}. Without this pulse, approximately 20\% of atoms in $\ket{r}$ return to $\ket{g}$ before detection as they are trapped \cite{Wilson2019}.

% The control beam is applied simultaneously with $\Omega_r$, and again with maximum power at the end of 

% To control the Rydberg excitations, the control beam is turned on with varying detunings and powers during the Rydberg pulse. In all experiments, the control beam is pulsed on again at maximum power after the Rydberg beam is turned off, to ionize any atoms remaining in $\ket{r}$ for the purpose of state detection \cite{Madjarov2020}. Without this pulse, approximately 20\% of atoms in $\ket{r}$ return to $\ket{g}$ before the final fluorescence image is captured.

The essential result is illustrated in Fig. \ref{fig:fig1}e. With the control beam turned off, we observe Rabi oscillations between $\ket{g}$ and $\ket{r}$ with a high visibility. Applying the control beam with successively higher intensities, these oscillations become damped and eventually cease, with the atoms remaining in the initial state $\ket{g}$. The control light is detuned by $\Delta=-5$ GHz from the $\ket{r} \rightarrow \ket{r'}$ transition. They decay from $\ket{r'}$ is predominantly via autoionization, which leads to atom loss. Therefore, the high survival probability in $\ket{g}$ directly illustrates that the rate of these scattering events is small. We note that only modest control beam power is required: the highest $I_c$, 600 W/cm$^2$, is equivalent to only 1.3 $\mu$W in a diffraction-limited spot with $w_0 = \lambda$.

%
 
% , so the suppression of coherent dynamics is best understood as a quantum Zeno effect \cite{itano1990,misra1977}

% JDT edited to here 7/18pm

\begin{figure}[tp]
\centering
\includegraphics[width=1\linewidth]{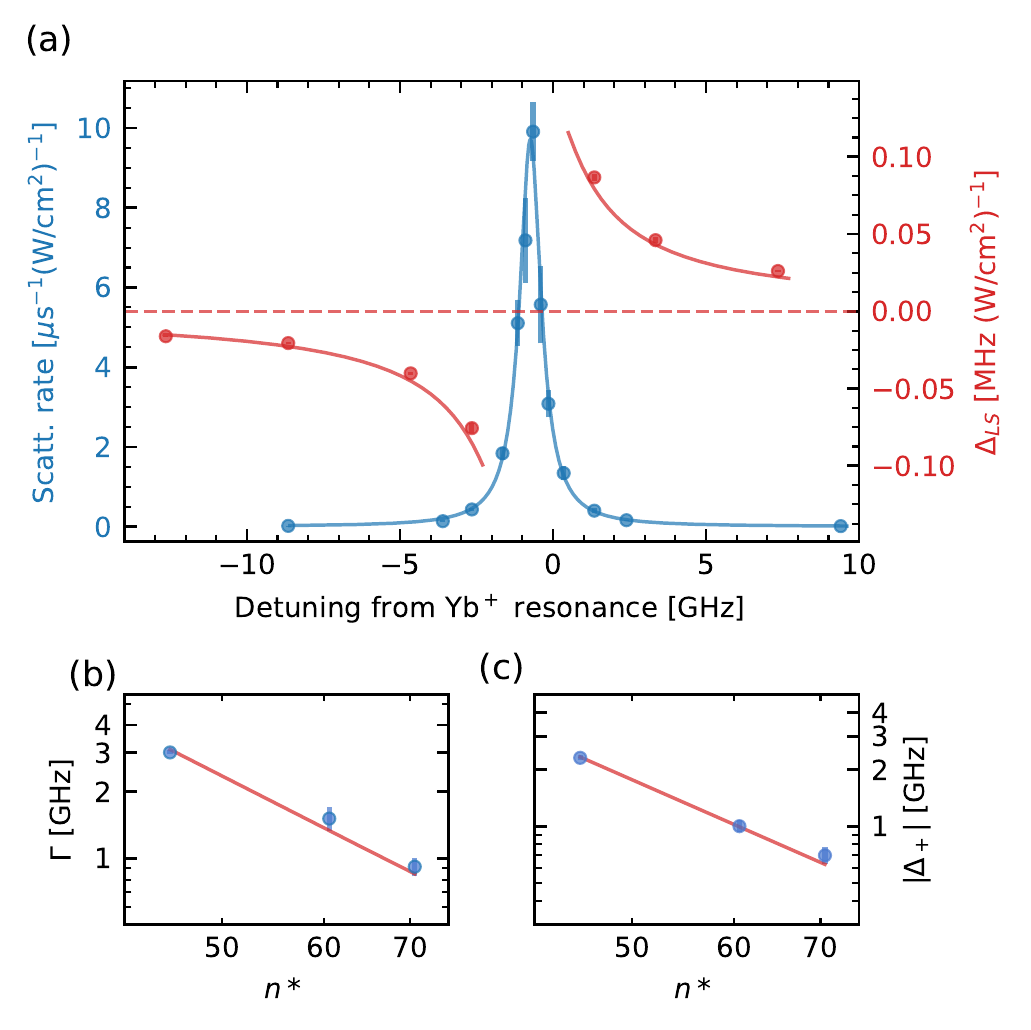}
\vspace{-3mm}
\caption{\label{fig:fig2}(a) $\Delta_{LS}$ (red) and $\Gamma_{LS}$ (blue) for the $6s75s$ $^3$S$_1$ state, as a function of the control beam detuning from the Yb$^+$ $6s \rightarrow 6p_{1/2}$ transition frequency, $f_+$. The lines are fits to the expressions in the text. (b) The extracted linewidth, $\Gamma$ and (c) center frequency offset $\Delta_+$ as a function of $n^*=n-\delta$ with $\delta=4.439$ \cite{Wilson2019}. A fit to $(n^*)^{-3}$ is included from which we extract the following: $\Gamma(n^*)=2\pi\times2.9\times10^{14}\text{ s}^{-1}/n^{*3}$ and $|\Delta_+(n^*)|=2.2\times10^{14}\text{ Hz}/n^{*3}$. }
\vspace{0mm}
\end{figure}

To better understand and quantitatively model the properties of the $\ket{r} \rightarrow \ket{r'}$ transition, we measure the light shift and scattering rate of $\ket{r}$ in the presence of the 369 nm control beam with variable detuning from the Yb$^+$ resonance at $f_+ = 811.29150(40)$ THz (Fig. \ref{fig:fig2}a) \cite{zalivako2019}. The light shift, $\Delta_{LS}$, is extracted from the resonance shift of the $\ket{g} \rightarrow \ket{r}$ transition, while the scattering rate, $\Gamma_{LS}$, is measured from the lifetime of $\ket{r}$ \cite{Wilson2019}, assuming that most scattering events result in autoionization of the atom. These measurements have been performed over a range of powers, with linear scaling observed in all cases.

% We also note that the peak scattering rate of $9.9(7)$ $\mu\textrm{s}^{-1}$/(W/cm$^2$) is more than 100 times larger than what was reported in Ref. \cite{Madjarov2020} for the Sr $5s61s \rightarrow 5p_{3/2}ns$ transition.

The data is well-described by a two-level system model, $\Delta_{LS} = \Omega_c^2\Delta/(4 \Delta^2 + \Gamma^2)$ and $\Gamma_{LS} = \Gamma \Omega_c^2/(4 \Delta^2 + \Gamma^2)$ \cite{grimm2000}. From a fit to the scattering rate, $\Gamma_{LS}$, we extract the parameters $\Delta = f - (f_+ + \Delta_+)$, $\Gamma = 2\pi\times0.92(3)$ GHz and a dipole moment $d = \hbar \Omega_c/E_c = 1.46(2) e a_0$ (where $E_c$ is the electric field strength of the control beam). The transition is centered at a frequency $\Delta_+ = -0.73(7)$ GHz below the Yb$^+$ transition frequency. The dipole moment is also consistent with that of the bare Yb$^+$ $6p_{1/2}$ transition \cite{olmschenk2009}, taking into account a Clebsch-Gordan coefficient for the $\sigma^{+/-}$ polarization of the laser. The same model describes the light shift $\Delta_{LS}$, although the magnitude of the light shift is approximately 30\% smaller than would be predicted from the parameters above. We have repeated these measurements at $n=50$ and 65 to understand the scaling properties of these parameters. Both $|\Delta_+|$ and $\Gamma$ scale as $1/n^{*3}$ as expected (Fig. \ref{fig:fig2}b and c) \cite{gallagher1987}.

\begin{figure}[htp]
\centering
\includegraphics[width=1\linewidth]{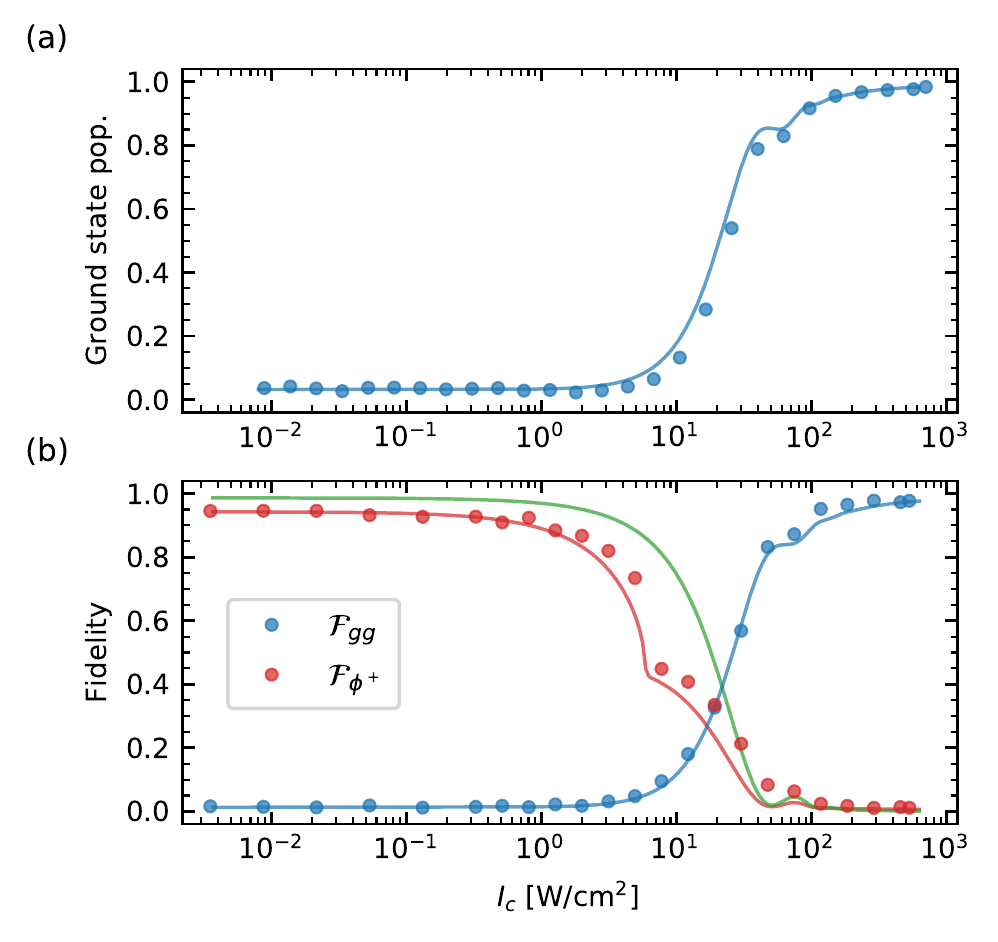}
\vspace{-5mm}
\caption{\label{fig:fig3} (a) Probability to find an atom in $\ket{g}$ after an attempted $\pi$ pulse to $\ket{r}$, for varying control intensities $I_c$ with $\Delta = -5$ GHz. At high intensities, the population transfer is strongly suppressed and 98.7(2)\% of the atoms remain in $\ket{g}$. (b) Final state fidelity with $\ket{\phi^+}$ (red points: experimental lower bound) and $\ket{gg}$ (blue) after a $\pi$ pulse in a dimerized array with a strong intra-dimer Rydberg blockade. The control beam switches the final state from having $>$ 0.948(12) fidelity with $\ket{\phi^+}$, to 0.978(3) fidelity with $\ket{gg}$. The curves show simulations of the same quantities using parameters from Fig. \ref{fig:fig2}a and adding phenomenological dephasing and detection errors. The green curve shows the exact fidelity $\mathcal{F}_{\phi^+}$, while the red curve shows the experimentally measurable lower bound on $\mathcal{F}_{\phi^+}$.}
\vspace{0mm}
\end{figure}

We now study the effectiveness of the control beam at modulating Rydberg excitation induced by $\Omega_r$ in the case of isolated, non-interacting atoms (Fig. \ref{fig:fig3}a). To do this, we apply $\Omega_r$ for a time $t_g=\pi/\Omega_r\approx700$ ns ($\pi$-pulse), while varying the control beam intensity $I_c$. The control beam detuning is fixed at $\Delta = -5$ GHz. In this experiment, the atoms are spaced by $d=21\,\mu$m, such that the van der Waals interaction is negligible compared to $\Omega_r$. In the absence of the control beam, around 97\% of the ground state population is transferred to $\ket{r}$, limited by residual technical noise, Doppler shifts and photon scattering. However, at the maximum $I_c$, the Rydberg excitation is suppressed and the atoms remain in $\ket{g}$ with 98.7(2)\% probability. The data is in excellent agreement with a numerical simulation based on parameters extracted from Fig. \ref{fig:fig2}a \cite{supplement}.

Next, we demonstrate control over coherent excitations in the Rydberg blockade regime using a dimerized array (Fig. \ref{fig:fig3}b). The intra-dimer spacing is 3.15 $\mu$m, for which we estimate a blockade strength $V \gg 100$ MHz \footnote{We extract a scaled $C_6$ coefficient of $C_6n^{-11}\approx-21^{+9}_{-7} (a.u.)$ at $n=50$ by determining the onset of blockaded Rabi oscillations vs. the separation between atoms. The error is due to the uncertainty of our imaging system magnification}. We apply $\Omega_r$ for a time $t_g = \pi/(\sqrt{2}\Omega_r) \approx 500$ ns, which drives a $\pi$ pulse on the blockaded transition $\ket{gg} \rightarrow \ket{\phi^+} = (\ket{gr}+\ket{rg})/\sqrt{2}$. When $I_c$ is small, we extract a lower bound of the state fidelity (with respect to the Bell state $\ket{\phi^+}$) of $\mathcal{F}_{\phi^+} >  0.948(12)$, based on a lower bound on the state purity after a $2t_g$ pulse  \cite{Madjarov2020, supplement}. However, at the maximum $I_c$, the oscillations are highly suppressed and the fidelity with the starting state is $\mathcal{F}_{gg} = 0.978(3)$.

%In the absence of 369 nm light, a Rydberg pulse of duration $\pi/(\sqrt{2}\Omega)$ excites the state $\ket{\phi^+} = (\ket{gr}+\ket{rg})/\sqrt{2}$ with fidelity $\mathcal{F_\phi} > \mathcal{F^>_\phi} = 0.948(X)$. This lower bound is determined from a lower bound on the state purity using the method of Madjarov et al. \cite{Madjarov2020}. At the highest 369 nm intensity, the oscillations are highly suppressed and the population remains in $\ket{gg}$ with 98\% probability. $P_{gg}$ is in excellent agreement with a theoretical model. The Bell state fidelity lower bound $\mathcal{F^>_\phi}$ is compared to a model for the true fidelity $\mathcal{F_\phi}$. It follows a similar trend, but sits below $\mathcal{F_\phi}$, which is expected since the lower bound is not tight when the fidelity is not close to one.
% JDT: Think about whether this claim about the lower bound is actually true.

\begin{figure}[tp]
\centering
\includegraphics[width=1\linewidth]{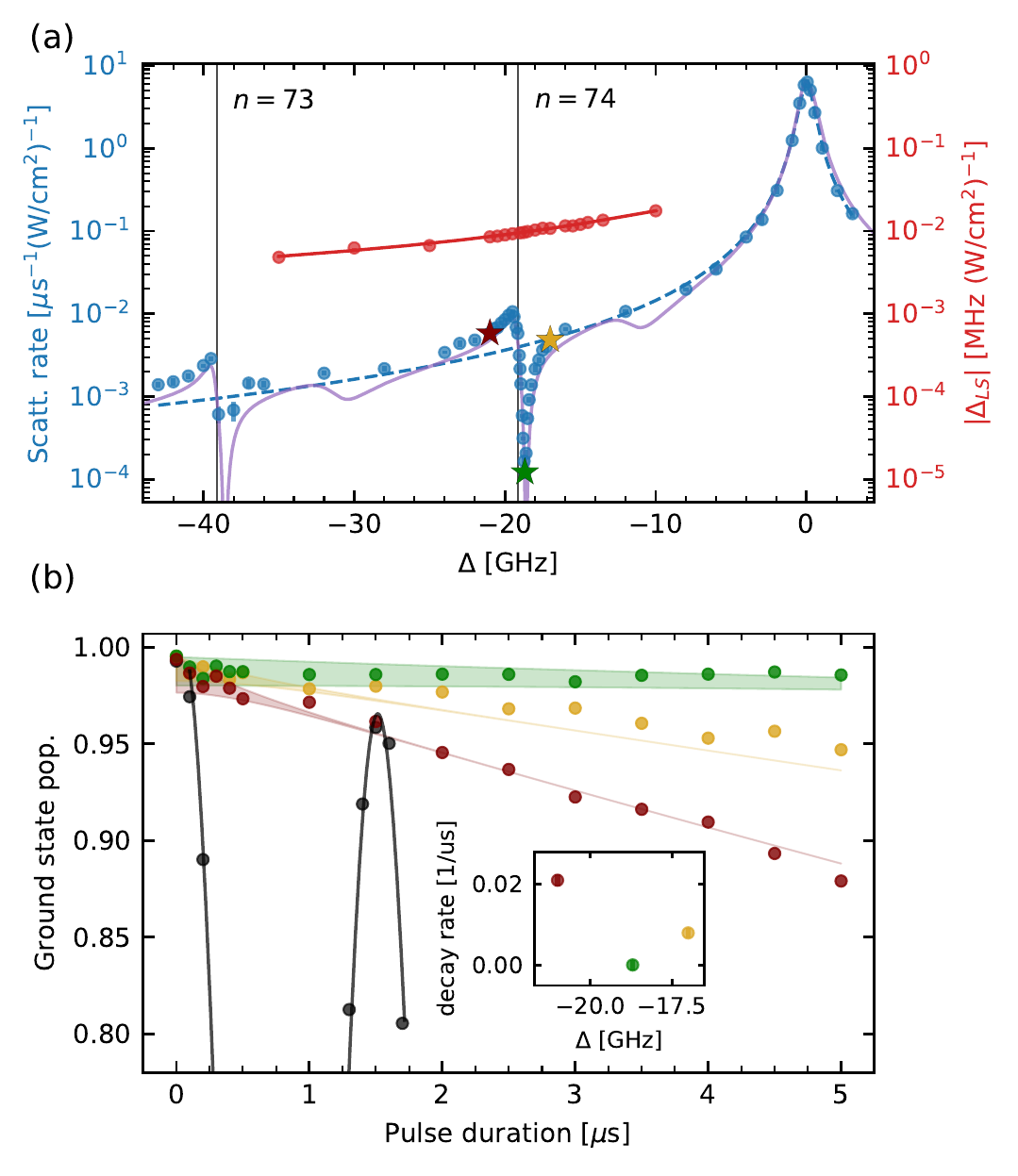}
\vspace{-5mm}
\caption{\label{fig:fig4}(a) $|\Delta_{LS}|$ (red) and $\Gamma_{LS}$ (blue) for the $6s75s$ $^3$S$_1$ state, as in Fig. \ref{fig:fig2} but over a broader range of detunings with a fit of $\Gamma_{LS}$ (blue dashed line) and a MQDT model (purple line) of the autoionization spectrum described in the supplementary information \cite{supplement}. The positions of the $6p_{1/2}ns$ states (black lines) and the detuning that minimizes the scattering rate $\Delta_{min} = -18.7$ GHz (green star) are indicated. (b) Rabi oscillations to $\ket{r}$ (black, $I_c=0$) are strongly suppressed by applying the control field at detunings near $\Delta_{min}$ [colors correspond to starred points in panel (a)] and show good agreement with simulation (solid lines). We find that population decay (inset) from autoionization is suppressed to within our experimental resolution when $\Delta=\Delta_{min}$.}
\vspace{0mm}
\end{figure}

Lastly, we study the the scattering rate and light shift over a broader range of control beam detunings. Near the transition to other Rydberg states, such as $6s75s \rightarrow 6p_{1/2}74s$ ($\Delta \approx -19$ GHz below the transition to $6p_{1/2}75s$), the scattering rate shows sharp, Fano-like features (Fig. \ref{fig:fig4}a) known as ``shake-up" resonances \cite{gallagher1987}. These features can be reproduced with a MQDT model \cite{supplement}, and arise from zeros in the overlap integral of the Rydberg electron wavefunction in the initial and final states \cite{tran1982}. Experimentally, we observe a dip in the scattering rate that is a factor of 34 below the value predicted by the two-level model for $\Gamma_{LS}$.

% $0^{+0.06}_{-0}$\% per $\pi$-pulse

Importantly, the light shift does not show any irregular behavior near the shake-up resonance. Therefore, operating the control laser at a detuning corresponding to the scattering minimum ($\Delta_{min}$) is beneficial, reducing $\epsilon_{sc}$ or allowing the same total addressing error $\epsilon_{rot} + \epsilon_{sc}$ to be realized with less power. We experimentally demonstrate the reduction in $\epsilon_{sc}$ by simultaneously applying $\Omega_r$ and the control light for a long duration ($t \approx 7 \pi/\Omega_r$), at three detunings near $\Delta_{min}$ (Fig. \ref{fig:fig4}b). The light shift ($\Delta_{LS} \approx 5.8-4.6$ MHz) suppresses coherent oscillations to $\sim$ 1\% for each detuning. However, there is gradual population loss from $\epsilon_{sc}$ that is linear in $\Gamma_{LS}$, and we find that this loss vanishes, within our experimental resolution at $\Delta_{min}$. The behavior agrees with a master equation simulation, where the shake-up resonance is incorporated as a reduction in $\Gamma$ to match the experimentally measured $\Gamma_{LS}$ at each of the starred detunings in Fig. \ref{fig:fig4}a.

In conclusion, we have demonstrated that a light shift of the Rydberg states using a core electron transition can be used to control Rydberg excitations in an optical tweezer array. In contrast to applying a light shift on the ground state, shifting the Rydberg state has reduced laser power requirements to achieve a given error rate from photon scattering or off-target excitation. This demonstration was carried out with a global control laser, however, the main application of this result will be with locally addressed control fields, and using the $^{171}$Yb isotope with qubits encoded in the nuclear spin sublevels of the $J=0$ $^1$S$_0$ or $^3$P$_0$ states. The latter approach has the additional benefit that the qubit states experience negligible differential light shift from the control field, significantly relaxing the technical demands for local control.

We acknowledge helpful conversations with Dolev Bluvstein. A.P.B., S.M., S.S., J.W. and J.D.T. were supported by ARO PECASE (W911NF-18-10215), ONR (N00014-20-1-2426), DARPA ONISQ (W911NF-20-10021) and the Sloan Foundation. S.S. was additionally supported by an ARO QuaGCR fellowship. M.A.A. and C. H. G. are supported in part by the AFOSR-MURI, grant number FA9550-20-1-0323.

\vspace{5mm}

\bibliography{autoionize.bib}
\bibliographystyle{apsrev4-2}
\end{document}

\title{Supplementary Information for ``Controlling Rydberg excitations using ion core transitions in alkaline earth atom tweezer arrays"}
%\title{Coherent control of Rydberg excitations using ion core transitions in an alkaline earth atom}
\author{Alex P. Burgers}
\affiliation{Princeton University, Department of Electrical and Computer Engineering, Princeton, New Jersey 08544}
\author{Shuo Ma}
\affiliation{Princeton University, Department of Electrical and Computer Engineering, Princeton, New Jersey 08544}
\affiliation{Princeton University, Department of Physics, Princeton, New Jersey 08544}
\author{Sam Saskin}
\affiliation{Princeton University, Department of Electrical and Computer Engineering, Princeton, New Jersey 08544}
\affiliation{Princeton University, Department of Physics, Princeton, New Jersey 08544}
\author{Jack Wilson}
\affiliation{Princeton University, Department of Electrical and Computer Engineering, Princeton, New Jersey 08544}
\author{Miguel A. Alarc\'on}
\author{Chris H. Greene}
\affiliation{Purdue University, Department of Physics and Astronomy, West Lafayette, Indiana 47907}
\affiliation{Purdue University, Purdue Quantum Science and Engineering Institute, West Lafayette, Indiana 47907}
\author{Jeff D. Thompson}
\affiliation{Princeton University, Department of Electrical and Computer Engineering, Princeton, New Jersey 08544}
\email{jdthompson@princeton.edu}
% \date{May 2020}
\maketitle

\section{MQDT}
In order to describe the autoionization process we propose a five channel model that aims to fit the spectrum by finding the appropriate multichannel quantum defect theory (MQDT) parameters. The control laser couples to odd parity $6p_{1/2}nl$ states with both $J'=0$ and $J'=1$. Under the assumption that photoionization occurs via an isolated core excitation (ICE) process, the odd $J'=2$ symmetry is not accessible for final Rydberg states attached to the $6p_{1/2}$ threshold. Omitting the open $f$ shell channels that are suspected to be irrelevant, there are in total $18$ channels, from which $5$ correspond to $J'=0$ and $13$ correspond to $J'=1$.

The simplest semi-realistic model includes only a single open channel (as proposed for Ba in Ref. \cite{Tran1984}), and only the $6p_{1/2}$ closed channels, which results in a total of $5$ channels: two for $J'=0$ and three for $J'=1$ (Table \ref{tab:tabChanJ}). The latter restriction is justified by the form of the dipole matrix elements in the ICE, which involves the overlap between the radial wavefunctions in the closed channels.

\begin{table}[htb]
    \centering\begin{tabular}{|c|c|}
        \hline \multicolumn{2}{|c|} {$\mathrm{J}=0$} \\
        \hline $j=1$ & $6 s_{1/2} \epsilon p_{1/2}$\\
        \hline $j=2$ & $6 p_{1/2}  n s_{1/2}$\\
        \hline
    \end{tabular}
    \centering\begin{tabular}{|c|c|}
        \hline \multicolumn{2}{|c|} {$\mathrm{J}=1$} \\
        \hline $j=1$ & $6 s_{1/2} \epsilon p_{1/2}$\\
        \hline $j=2$ & $6 p_{1/2}  n s_{1/2}$ \\
        \hline $j=3$ & $6 p_{1/2}  n d_{3/2}$ \\
        \hline
    \end{tabular}
    \caption{Five channel model with a single continuum.}
    \label{tab:tabChanJ}
\end{table}
\begin{table*}[tp]
    \centering
    \begin{tabular}{|c| c| c|}
        \hline
        $\mu^0_{1,1}$ &$\mu^0_{1,2}$ &$\mu^0_{2,2}$  \\[2mm]
        \hline 
        \hline
        $8.58074\times10^{-3}$ &  $1.71383\times10^{-1}$ &  $-4.83877\times10^{-1}$  \\
        \hline
    \end{tabular}\\
    \bigskip
    \bigskip
    \begin{tabular}{|c| c |c |c |c |c|}
        \hline
        $\mu^1_{1,1}$ &$\mu^1_{1,2}$ &$\mu^1_{1,3}$ &$\mu^1_{2,2}$ &$\mu^1_{2,3}$ &$\mu^1_{3,3}$ \\[2mm]
        \hline 
        \hline
         $3.68692\times 10^{-2} $ & $-1.37765$& $4.42814\times 10^{-2}$ & $-1.35495\times 10^{-2} $ & $ -7.41744\times 10^{-1} $ & $ 1.02353\times 10^{-2} $ \\
        \hline
    \end{tabular}
    \caption{$\mu$ matrix elements corresponding to the fit in Fig. 4 of the main text. }
    \label{tab:params}
\end{table*}

The matrix $K$ describing the short-range behavior is parameterized by the $\mu$ matrix \cite{Du1986}, which we fit to the experimental data. This matrix is connected to the $K$ matrix by $K=\tan(\pi \mu)$ where the $\tan$ is understood to be a matrix function.

To compare to the data, we compute the photoionization rate, $R$ as a function of the energy of the autoionizing atomic state $E=E_{75}+\omega$, where $E_{75} = 50421.0303$ cm$^{-1}$ is the energy of the $6s75s$ $^3$S$_1$ state \cite{Wilson2019}, and $\omega$ is the control laser frequency. Adopting the notation of Ref. \cite{aymar1996} for the outgoing wave function of the autoionizing state, and the approximate overlap of the closed channel functions \cite{Bhatti1981}, we find:
\begin{equation}
    \begin{split}
    \label{eq:cross}
    R(E) = &\frac{\pi}{2} \mathcal{E}_o^2 \mathcal{D}^2_{6p_{1/2},6s_{1/2}} \left( \frac{2 \sin\left[\pi(\nu(E)-\nu_0)\right]\nu(E)^2 \nu_0^2}{\sqrt{6} \nu_0^{3/2}\pi(\nu(E)^2-\nu_0^2)} \right)^2  \left(\left|Z^{J=1}_{21}(E)\right|^2+\left|Z^{J=0}_{21}(E)\right|^2\right)
    \end{split}
\end{equation}
Here $\nu_0=70.561$ is the effective quantum number of the $6s75s $ $^3$S$_1$ state with respect to the $6s_{1/2}$ threshold \cite{Wilson2019}, and $\nu(E)= \sqrt{R_{\text{Yb}}/(I_{6p_{1/2}}-E)}$ is the effective quantum number of the final autoionizing state with respect to the $6p_{1/2}$ threshold, $I_{6p_{1/2}}$, with the Yb Rydberg constant $R_{\text{Yb}}=R\times m_{\text{Yb}}/(m_{\text{Yb}}+m_e)$. $Z_{21}(E)$ is the coefficient of the closed channel function in the incoming wave boundary condition wave function as described in Ref. \cite{aymar1996}, and $\mathcal{D}_{6p_{1/2},6s_{1/2}} = 2.6829e a_0$ is the reduced matrix element of the electric dipole operator in the core states \cite{Safronova2009}. The $\mu$ matrix that produces the best agreeemnt with the experimental photoionization spectrum (Fig. 4a) is given in Table \ref{tab:params}.

The counter-intuitive structure of this spectrum is worth some explanation. Since the electric dipole matrix element for an ICE is proportional to the overlap between closed channel functions, and in this limited MQDT model all the closed channels are attached to the same threshold, the ionization rates vanish exactly when $\nu_0-\nu(E)$ is an integer different from zero. This exact vanishing will not happen if closed channels attached to the $6p_{3/2}$ channels are included, which may be the origin of the finite scattering rate at $\Delta_{min}$ observed in the experiment. This can also be modified by considering an initial wave function that includes amplitude in channels other than one attached to the $6s_{1/2}$ threshold. However, a complete MQDT model for the $^3$S$_1$ series in $^{174}$Yb does not exist as this series was only recently observed for the first time \cite{Wilson2019}.

The five-channel model has more parameters than can be accurately constrained by the experimental data. In order to fit both the satellite lines and the main feature, it has proved necessary for the $J'=0$ and $J'=1$ $ns$ resonances to be close in energy and overlapping, since the main feature at zero detuning does not suggest a composite peak. However, their relative energy ordering and widths are not strongly constrained by the current experimental data, which does not distinguish these symmetries. Similarly, the small outlying feature in the fit at detuning $\Delta\approx -11$ GHz and $\Delta\approx -31$ GHz comes from the zero of a Fano resonance \cite{Fano1961} in the $J'=1$ states. This feature is not observed in the sparsely sampled data in this detuning range, and could in fact lie elsewhere.

\begin{figure}[htp]
\centering
\includegraphics[width=0.6\linewidth]{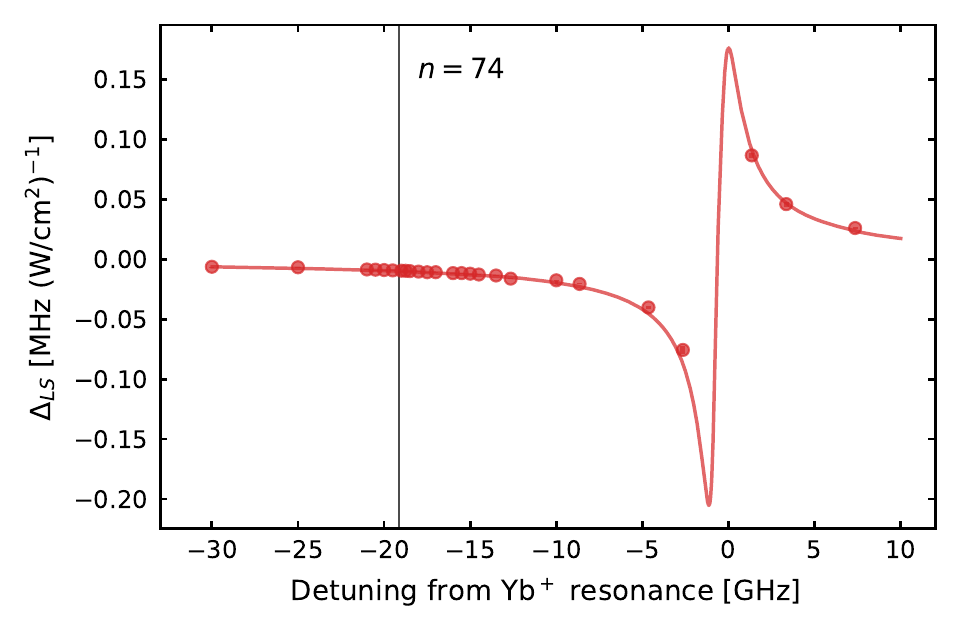}
\vspace{0cm}
\caption{\label{fig:supp1} Scaled light shift curve using the MQDT model parameters and Eqn. \ref{eq:lightshift} with data from the main text. Importantly, the MQDT model for the light shift does not predict a feature near the satellite lines (n=74 indicated with a black line).}
\end{figure}

We can also compute the light shift predited by the MQDT model. Adapting the formula of Ref. \cite{Shore1990} to the autoionizing states, we find the following energy shift:

\begin{widetext}
\begin{equation}\label{eq:lightshift}
   \begin{aligned}
        \Delta E_{75} = -\frac{1}{4}\mathcal{E}_o^2 \mathcal{D}_{6p_{1/2}, 6s_{1/2}}^2 
        &\left( P\int \left( \frac{2 \sin \left[\pi (\nu(E)-\nu_0)\right] \nu(E)^2 \nu_0^2}{\sqrt{6} \nu_0^{3/2} \pi(\nu(E)^2-\nu_0^2)} \right)^2 \left( |Z^{J=1}_{21}(E)|^2+|Z^{J=0}_{21}(E)|^2\right) \right.  \left.
         \frac{1}{E-E_{75}-\omega} dE \right.\\[2mm]
        & \left. \left. \right. \right. + \left. i \pi \left. \left( \frac{2 \sin \left[\pi (\nu-\nu_0)\right] \nu^2 \nu_0^2}{\sqrt{6} \nu_0^{3/2} \pi(\nu^2-\nu_0^2)} \right)^2 \left( |Z^{J=1}_{21}|^2+|Z^{J=0}_{21}|^2\right) \right|_{E=E_{75}+\omega} \right),
   \end{aligned}
\end{equation}   
\end{widetext}
where $P$ indicates the Cauchy principal value of the integral. The real part of $\Delta E_{75}$ gives the light shift, while the imaginary part reproduces the photoionization rate in agreement with Eq. \eqref{eq:cross}, $R=-2\text{Im}(\Delta E_{75})$. In Fig. \ref{fig:supp1}, we plot the real part along with the light shift data from Figs. 2a and 4a from the main text. The predicted light shift from the MQDT model is scaled by 0.7 to account for the 30\% discrepancy between the predicted light shift from the scattering rate model and the data, as was noted in the main text. The absence of a feature in the light shift near the satellite lines reflects the fact that the integral in Eq. \ref{eq:lightshift} derives most of its weight from the main resonance near zero detuning.

\section{Bell state fidelity measurement}

The Bell state fidelity $\mathcal{F}_{\phi_+}$ is not directly accessible in the experiment without a local probe or the ability to perform single-qubit rotations on the $\ket{g}-\ket{r}$ subspace without the influence of the Rydberg blockade. Therefore, we quantify the fidelity with a lower bound using the method from Ref. \cite{Madjarov2020}, which is based on the state purity extracted from the populations at a time $2t_g$:
\begin{equation}
\begin{split}
    \mathcal{F}_{\phi_+}(t_g) &> \frac{1}{2}\left[\rho_{gr,gr}(t_g) + \rho_{rg,rg}(t_g)\right] +
    \sqrt{\textrm{max}(0,(\sum_i \rho_{i,i}(2 t_g)^2-1)/2 + \rho_{gr,gr}(t_g) \rho_{rg,rg}(t_g))}
\end{split}
\label{eq:fidelity}
\end{equation}
We use the two-atom basis populations (shown in Table \ref{tab:pop}) averaged over the smallest three values of $I_c$ to calculate the fidelity quoted in the text, $\mathcal{F}_{\phi+}>0.948(12)$. We note that these populations correspond to raw measurement outcomes, and are not corrected for any state preparation or measurement errors.

\begin{table}[]
    \centering
    \begin{tabular}{| c | c | c |}
    \hline
           & $t_g$ & $2t_g$ \\
           \hline
          $\rho_{gr,gr}$ & 0.487(8) & 0.013(2) \\
          $\rho_{rg,rg}$ & 0.494(8) & 0.013(2) \\
          $\rho_{rr,rr}$ & 0.005(1) & 0.006(1) \\
          $\rho_{gg,gg}$ & 0.014(2) & 0.968(3) \\
          \hline
    \end{tabular}
    \caption{Two-atom basis populations for the lowest three values of $I_c$ at $t_g$ and $2t_g$ used to calculate the fidelity from Eqn. \ref{eq:fidelity}}
    \label{tab:pop}
\end{table}

\section{Simulations}
Here we provide details of the single and two-atom simulations used to model the Rydberg excitation dynamics in the presence of the control field, as presented in Fig. 3 of the main text.

\subsubsection{Isolated atoms}
For isolated, non-interacting atoms, we solve the master equation for the following Hamiltonian:
\begin{equation}
    H_{1} = \frac{\Omega_{r}}{2}(\sigma^{+}_{r,g} + \sigma^{-}_{r,g}) + \frac{\Omega_{c}}{2}(\sigma^{+}_{r,r'} + \sigma^{-}_{r,r'}) - \Delta P_{r'}
\end{equation}
We include an additional state, $\ket{d}$, in the Hilbert space to capture decay from the autoionizing state $\ket{r'}$, governed by the collapse operator $c_{1} = \sqrt{\Gamma}\sigma^{-}_{r',d}$. We also include an additional collapse operator, $c_{2} = \sqrt{\gamma_{\text{r}}}\sigma^{z}_{g,r}$ to incorporate the observed dephasing of the Rabi oscillations at a phenomenological level. The dephasing rate, $\gamma_{\text{r}}$, is chosen to match the simulated ground state population after a $\pi$-pulse, $P_g(\pi/\Omega_r)$, to the experimental value when $I_c = 0$. We then compute $P_g(\pi/\Omega_r)$ as a function of $I_c$ to generate the curve in Fig. 3a.

To facilitate the comparison with the experiment, the simulated $P_g$ is multiplied by the ground state detection fidelity $F_g = 0.994$, extracted from an independent measurement with no Rydberg excitation.

\subsubsection{Two atom blockade}
To model the Rydberg excitation dynamics in the presence of interactions, we extend our previous model into the following two-atom Hamiltonian:
\begin{equation}
    H_{2} = H_{1} \otimes  \mathbb{1} + \mathbb{1} \otimes H_{1} + U_{\text{int}}P_{r} \otimes P_{r}
\end{equation}
We now include four collapse operators: $c_{1} = \sqrt{\Gamma}\sigma^{-}_{r',d}\otimes  \mathbb{1}$, $c_{2} = \sqrt{\Gamma}\mathbb{1} \otimes \sigma^{-}_{r',d}$, $c_{3} = \sqrt{\gamma_r}\sigma^{z}_{g,r} \otimes \mathbb{1}$, $c_{4} = \sqrt{\gamma_r}\mathbb{1} \otimes \sigma^{z}_{g,r}$. In order to account for imperfect ground state detection we transform the simulated populations using the following matrix:

\begin{equation}
    T_{(\text{sim} \rightarrow \text{exp})} =
    \begin{pmatrix}
    F_{g}^{2} & 0 & 0 & 0 \\
    F_{g}(1-F_{g}) & F_{g} & 0 & 0 \\
    F_{g}(1-F_{g}) & 0 & F_{g} & 0 \\
    (1-F_{g})^{2} & 1-F_{g} & 1-F_{g} & 1
    \end{pmatrix}
\end{equation}
We directly simulate the ground state probability $P_{gg}(t_g)$ and the fidelity of the Bell state $\mathcal{F}_{\phi_+}=\Tr\left[\rho(t_g) \ket{\phi_+}\bra{\phi_+}\right]$ after a pulse of duration $t_g = \pi/\sqrt{2}\Omega_r$, for various $I_c$ (here, $\rho(t)$ is the two-atom density matrix at time $t$). We also compute the fidelity lower bound from Eq. \eqref{eq:fidelity} and plot this in Fig. 3a. The good agreement with the measurement suggests that the observed infidelity is significantly limited by the lack of a direct entanglement probe.
 
%We simulate $P_{gg}(t_g = \pi/(\sqrt{2}\Omega_{r}))$ and extract a bell-state fidelity lower bound (method from Ref. \cite{Madjarov2020}) from the simulated populations $P^{\text{ sim}}_{gr}(t_g)$, $P^{\text{ sim}}_{rg}(t_g)$, $P^{\text{ sim}}_{gg}(t_g)$, $P^{\text{ sim}}_{gr}(t_g)$, $P^{\text{ sim}}_{rg}(t_g)$, $P^{\text{ sim}}_{rr}(t_g)$  as a function of the control beam Rabi frequency, $\Omega_{c}$.

Lastly, we note that these simulations treat the $\ket{r}\rightarrow\ket{r'}$ transition as a two-level system. However, as discussed above, the observed light shift and scattering rate on this transition are not self-consistently described by a single linewidth and dipole moment, with the apparent strength of the light shift being smaller by a factor of 0.7. To incorporate this into the simulation, we increase the linewidth by a factor of $1/0.7$, and scale the control intensity by $0.7$. For values of $\Delta \gg \Gamma$, this accurately reproduces the experimentally measured $\Delta_{LS}$ and $\Gamma_{LS}$. Without this modification, the agreement between the simulation and experiment in Fig. 3b is considerably worse.

\bibliography{autoionize.bib}
\bibliographystyle{apsrev4-2}